\magnification \magstep1
\raggedbottom
\openup 2\jot
\voffset6truemm
\leftline {\bf EXACT SOLUTIONS OF THE SCHR\"{O}DINGER EQUATION}
\leftline {\bf WITH INVERSE-POWER POTENTIAL}
\vskip 0.3cm
\noindent
{\bf Shi-Hai Dong}$^{1}$, {\bf Zhong-Qi Ma}$^{2,1}$ 
and {\bf Giampiero Esposito}$^{3,4}$
\vskip 0.3cm
\noindent
{\it ${ }^{1}$Institute of High Energy Physics, P. O. Box 918(4), 
Beijing 100039, People's Republic of China}
\vskip 0.3cm
\noindent
{\it ${ }^{2}$China Center for Advanced Science and Technology
(World Laboratory), P. O. Box 8730, Beijing 100080}
\vskip 0.3cm
\noindent
{\it ${ }^{3}$INFN, Sezione di Napoli, Mostra d'Oltremare
Padiglione 20, 80125 Napoli, Italy}
\vskip 0.3cm
\noindent
{\it ${ }^{4}$Dipartimento di
Scienze Fisiche, Complesso Universitario di Monte S. Angelo,
Via Cintia, Edificio G, 80126 Napoli, Italy}
\vskip 1cm
\noindent
The Schr\"{o}dinger equation for stationary states is studied in 
a central potential $V(r)$ proportional to $r^{-\beta}$ in an
arbitrary number of spatial dimensions. The presence of a single
term in the potential makes it impossible to use previous 
algorithms, which only work for quasi-exactly-solvable problems.
Nevertheless, the analysis of the stationary Schr\"{o}dinger
equation in the neighbourhood of the origin and of the point at
infinity is found to provide relevant information about the desired
solutions for all values of the radial coordinate. The original
eigenvalue equation is mapped into a differential equation with
milder singularities, and the role played by the particular case
$\beta=4$ is elucidated. In general, whenever the parameter $\beta$
is even and larger than $4$, a recursive algorithm for the 
evaluation of eigenfunctions is obtained.
Eventually, in the particular case of two spatial dimensions, the
exact form of the ground-state wave function is obtained for a
potential containing a finite number of inverse powers of $r$,
with the associated energy eigenvalue.
\vskip 100cm
\leftline {\bf 1. INTRODUCTION}
\vskip 0.3cm
\noindent
Many efforts have been produced in the literature over several
decades to study the stationary Schr\"{o}dinger equation in 
various dimensions with a central potential containing negative
powers of the radial coordinate [1--19]. Some relevant examples
are as follows. 
\vskip 0.3cm
\noindent
(i) When gaseous ions or electrons move through a gas whose
molecules are not too large, then the two interact according
to the law [3]
$$
V(r)=-{1\over 2}e^{2}\alpha r^{-4},
\eqno (1.1)
$$
where $e$ is the ionic charge, $\alpha$ the molecular polarizability,
and $r$ the distance between the ion and the molecule.
\vskip 0.3cm
\noindent
(ii) Interactions in one-electron atoms, muonic and hadronic and
Rydberg atoms; photodecay of excited states.
\vskip 0.3cm
\noindent
(iii) Repulsive singular potentials, for which the stationary
Schr\"{o}dinger equation has non-Fuchsian singularities at zero
and at infinity, may be sometimes expressed 
by a Laurent series [12--18]
$$
V(r)=\sum_{n=-\infty}^{\infty}a_{n}r^{n}
\eqno (1.2)
$$
in the annulus $r \in ]0,\infty[$.
\vskip 0.3cm
\noindent
(iv) The Dirac equation for a spin-${1\over 2}$ particle interacting
with scalar, electric and magnetic potentials leads to second-order
equations for spinor wave functions involving an effective potential
containing terms proportional to $r^{-2}, r^{-3}$ and $r^{-4}$. 
Such equations can be used to prove the existence of magnetic
resonances between massive and massless spin-${1\over 2}$ particles
with magnetic moments [19].

Major progress has been possible when the potential contains
a finite number of powers of $r$, e.g. $r^{-1}, r^{-2}, r^{-3}$
and $r^{-4}$, or $r^{2}, r^{-4}$ and $r^{-6}$ [7]. Such a property
is not a mathematical accident, because in many cases one only
succeeds in finding some eigenstates and eigenvalues for the
so-called quasi-exactly-solvable problems [20]. What happens is
that an algorithm can be found to evaluate some eigenfunctions
provided that a number of coefficients in the potential remain
non-vanishing and obey some restrictions, so that enough algebraic
equations are obtained for all unknown parameters in the ansatz.

The truly hard mathematical task, however, remains the one of
solving the stationary Schr\"{o}dinger equation with only one 
term in the potential, or for potentials admitting a Laurent
series expansion for $r \in ]0,\infty[$. The latter problem is
studied by several authors (see [12--18] and references therein)
and hence we focus on the former. When the stationary Schr\"{o}dinger
equation is studied for a central potential $U(r)$ in ${\bf R}^{q}$,
it reads [18]
$$
\left[{d^{2}\over dr^{2}}+{(q-1)\over r}{d\over dr}
+\kappa -{2m\over {\hbar}^{2}}U(r)
-{l(l+q-2)\over r^{2}}\right]\psi(r)=0,
\eqno (1.3)
$$
where $l(l+q-2)$ results from the action of the Laplace--Beltrami
operator on square-integrable functions on the $(q-1)$-sphere.
Equation (1.3) can be re-expressed in a form 
which does not involve first derivatives, i.e. [21]
$$
\left[{d^{2}\over dr^{2}}+\kappa-V(r)
-{(\lambda^{2}-{1\over 4})\over r^{2}}\right]y(r)=0,
\eqno (1.4)
$$
where we have defined [21] 
$$
\kappa \equiv {2mE \over \hbar^{2}},
\eqno (1.5)
$$
$$
V(r) \equiv {2m\over \hbar^{2}}U(r),
\eqno (1.6)
$$
$$
\lambda \equiv l+{1\over 2}(q-2),
\eqno (1.7)
$$
$$
y(r) \equiv r^{(q-1)\over 2} \psi(r),
\eqno (1.8)
$$
having denoted by $m$ the mass parameter, by $E$ the eigenvalues,
by $l$ the angular momentum quantum number 
and by $\psi(r)$ the radial part of the wave function.
The form (1.4) of the stationary Schr\"{o}dinger equation is the
most suitable for applications to scattering problems, because 
the operator in square brackets is an even function of
$\lambda$ [12, 21]. 

For a given form of inverse-power potential, Sec. 2 studies
the limiting form of Eq. (1.4) as $r \rightarrow 0$ and as
$r \rightarrow \infty$ to obtain a convenient factorization of
its solution for all values of $r$. This involves an unknown 
function $F$, and an algorithm for the evaluation of $F$ is
developed in Sec. 3, if the inverse power of $r$ in the
potential $V(r)$ is even. A class of ground-state wave functions
are evaluated in Sec. 4, and concluding remarks are presented
in Sec. 5.
\vskip 0.3cm
\leftline {\bf 2. PROPERTIES OF EIGENFUNCTIONS}
\vskip 0.3cm
\noindent
For the reasons described in the introduction, 
we are interested in a potential
$V$ having the form (hereafter $\beta > 2$)
$$
V(r) \equiv \alpha r^{-\beta},
\eqno (2.1) 
$$
where $\alpha$ and $\beta$ are some dimensionful and dimensionless
parameters, respectively. The first step in the attempt of solving
Eq. (1.4) is the search of regular solutions in the neighbourhood
of the origin. We therefore look for solutions $y(r)$ taking
the limiting form
$$
y(r)=r^{p}e^{-\gamma r^{-\delta}} \; {\rm as} \; 
r \rightarrow 0,
\eqno (2.2)
$$
where $p,\gamma$ and $\delta$ are some
parameters to be determined by consistency conditions, having taken
$\delta$ to be positive, as is suggested by known results for
some values of $\beta$, e.g. when $\beta=4$. Indeed, on inserting
the ansatz (2.2) into Eq. (1.4) one finds when $r \rightarrow 0$
the equation (for a fixed value of $\kappa$, which is therefore 
negligible with respect to inverse powers of $r$)
$$
{p(p-1)-(\lambda^{2}-{1\over 4}) \over r^{2}}
+{\gamma \delta (2p-\delta -1)\over r^{\delta +2}}
+{\gamma^{2}\delta^{2}\over r^{2\delta+2}}
-{\alpha \over r^{\beta}}=0.
\eqno (2.3)
$$
Since $r \rightarrow 0$ and $\delta >0$, this limiting form of
the equation reduces to
$$
{\gamma \delta (2p-\delta -1)\over r^{\delta+2}}
+{\gamma^{2}\delta^{2}\over r^{2\delta+2}}
-{\alpha \over r^{\beta}}=0.
\eqno (2.4)
$$
Of course, the term proportional to $r^{-2\delta-2}$ dominates
over the term proportional to $r^{-\delta -2}$ as 
$r \rightarrow 0$, and hence Eq. (2.4) can only be satisfied if
$$
\gamma^{2}\delta^{2}=\alpha ,
\eqno (2.5)
$$
$$
2\delta+2=\beta ,
\eqno (2.6a)
$$
which implies that the potential is repulsive (since
$\alpha >0$), and
$$
\delta={\beta \over 2}-1,
\eqno (2.6b)
$$
$$
\gamma={2\sqrt{\alpha}\over (\beta-2)}.
\eqno (2.7)
$$
So far, $p$ remains undetermined. Note, however, that the
coefficient $(2p-\delta-1)$ in Eq. (2.4) can be taken to
vanish. This assumption leads to
$$
\delta=2p-1,
\eqno (2.8)
$$
and by virtue of (2.6b) and (2.8) one finds $\beta=4p$. In
particular, if $p=1$, one recovers the familiar limiting form of
the solution when the repulsive potential is proportional
to $r^{-4}$:
$$
y(r)=r e^{-\sqrt{\alpha}r^{-1}} \; \;
{\rm as} \; r \rightarrow 0.
\eqno (2.9)
$$

In the neighbourhood of the point at infinity, solutions of
Eq. (1.4) behave as (hereafter $\varepsilon \equiv \pm 1$)
$$
y(r) \sim e^{i \varepsilon r \sqrt{\kappa}} \; \; as \;
r \rightarrow \infty,
\eqno (2.10)
$$
since only positive values of $E$ are compatible with a repulsive
potential. The ansatz for finding solutions of Eq. (1.4) 
{\it for all values} of $r$ when the potential (2.1) is considered
reads therefore
$$
y(r)=e^{-\gamma r^{-\delta}} \; 
e^{i\varepsilon r \sqrt{\kappa}} F(r),
\eqno (2.11)
$$
where the function $F$ interpolates in between the asymptotic 
regimes described by (2.2) and (2.10). By virtue of (1.4), (2.1) and
(2.11) the function $F$ should solve the differential equation
$$
\left[{d^{2}\over dr^{2}}+p(r){d\over dr}+q(r)\right]F(r)=0,
\eqno (2.12)
$$
where
$$
p(r) \equiv 2\gamma \delta r^{-\delta-1}
+2i \varepsilon \sqrt{\kappa},
\eqno (2.13)
$$
$$ \eqalignno{
q(r) & \equiv \gamma^{2}\delta^{2}r^{-2\delta-2}
-\gamma \delta (\delta+1)r^{-\delta-2}
+2i \varepsilon \gamma \delta \sqrt{\kappa}r^{-\delta-1} \cr
&-\alpha r^{-\beta}-{(\lambda^{2}-{1\over 4})\over r^{2}}.
&(2.14)\cr}
$$
\vskip 0.3cm 
\leftline {\bf 3. ALGORITHM FOR THE EVALUATION OF $F(r)$}
\vskip 0.3cm
\noindent
If we now assume that (2.5) and (2.6a) remain valid, the formulae
(2.13) and (2.14) reduce to
$$
p(r)=2\sqrt{\alpha}r^{-{\beta \over 2}}
+2i\varepsilon \sqrt{\kappa},
\eqno (3.1)
$$
$$
q(r)=-{\beta \over 2}\sqrt{\alpha}r^{-{\beta \over 2}-1}
+2i\varepsilon \sqrt{\alpha \kappa} \; r^{-{\beta \over 2}}
-{(\lambda^{2}-{1\over 4})\over r^{2}}.
\eqno (3.2)
$$
Since we are taking $\beta > 2$, such formulae lead to non-Fuchsian
singularities in Eq. (2.12), and hence we look for $F(r)$ in 
the form
$$
F(r)=r^{\omega} \sum_{s=-\infty}^{\infty} a_{s}r^{s}
\equiv r^{\omega} \sigma(r).
\eqno (3.3)
$$
It is crucial to allow for negative powers of $r$ in the solution,
which are associated to the non-Fuchsian nature of the singular
point at $r=0$. By virtue of (2.12) and (3.3), the series $\sigma$
obeys the second-order equation
$$
\left[{d^{2}\over dr^{2}}+\left({2\omega \over r}+p(r)\right)
{d\over dr}+\left({\omega (\omega-1)\over r^{2}}
+{\omega \over r}p(r)+q(r)\right)\right]\sigma(r)=0.
\eqno (3.4)
$$
In Eq. (3.4), the term ${\omega \over r}p(r)+q(r)$ has the
coefficient $\sqrt{\alpha} \left(2\omega-{\beta \over 2}\right)$
for $r^{-{\beta \over 2}-1}$. We set it to zero to get rid
of the dominant singularity at $r=0$, which implies
$$
\omega={\beta \over 4}.
\eqno (3.5)
$$
As a consistency check we point out that, if $\beta=4$, one recovers
the well known value $\omega=1$ (see (2.9)). For odd values of 
$\beta$, $\omega$ is therefore a polydromy parameter for the wave
function, as is familiar in singular potential scattering [13,17,18].

If Eq. (3.5) is taken to hold, and if $\beta$ is even, so that
${\beta \over 2}$ is an integer $b$, Eqs. (3.3) and (3.4) lead to
the following recurrence relation among coefficients of the
series $\sigma(r)$:
$$ \eqalignno{
\; & 2\sqrt{\alpha}(s+b+1)a_{s+b+1}
+2i\varepsilon \sqrt{\alpha \kappa}a_{s+b} \cr
&+\left[(s+2)(s+1)+{\beta^{2}\over 16}
+\beta \left({s\over 2}-{1\over 4}\right)
-\left(\lambda^{2}-{1\over 4}\right)\right]a_{s+2} \cr
&+\left[2i\varepsilon \sqrt{\kappa}(s+1)
+{i\over 2}\varepsilon \beta \sqrt{\kappa}\right]a_{s+1}=0,
&(3.6)\cr}
$$
for all integer values of $s$ from $-\infty$ through $+\infty$.
Such a formula is even more complicated than the recurrence
relation for coefficients of Mathieu functions [22].
Nevertheless, it provides a well defined rule for the evaluation
of $F(r)$ for given values of $\alpha$, when $\beta$ is
even. If $\beta$ is odd, we no longer get power series,
and hence we are unable to obtain a recursive algorithm.
\vskip 0.3cm
\leftline {\bf 4. A CLASS OF GROUND-STATE WAVE FUNCTIONS}
\vskip 0.3cm
\noindent
It may be of some interest to conclude our paper by considering a
different application of Eq. (1.4), i.e. the evaluation
of ground-state wave functions for potentials $V(r)$ containing 
more than one negative power of $r$. In such a case an algorithm 
for bound states, rather than scattering states,
can be found provided that the parameters in the potential obey
a set of suitable restrictions, and
we are now aiming to show how this can be obtained. For simplicity,
we set $q=2$ (bearing also in mind that two-dimensional models
are of interest), so that $\lambda=l$ (see (1.7)), and we consider
a potential in the form
$$
V(r)={A\over r^{4}}+{B\over r^{3}}+{C\over r^{2}}+{D\over r}.
\eqno (4.1)
$$
The term ${C\over r^{2}}$ might indeed be combined with the angular
momentum contribution ${(l^{2}-{1\over 4})\over r^{2}}$, but we prefer
to keep them distinct to emphasize their different origin. Our 
ground-state ansatz for $y(r)$ reads
$$
y(r)=\exp k(r),
\eqno (4.2)
$$
where
$$
k(r) \equiv {a\over r}+br+c \log(r) \; \; a < 0, \; \; b<0.
\eqno (4.3)
$$
The negative signs of the coefficients $a$ and $b$ are necessary 
to ensure square integrability at zero and at infinity, respectively.
The insertion of the ansatz (4.2) into Eq. (1.4) leads therefore
to the equation
$$
y''(r)-\Bigr[k''(r)+(k'(r))^{2}\Bigr]y(r)=0,
\eqno (4.4)
$$
where the prime denotes the derivative with respect to the variable
$r$. We now express $y''(r)$ from Eq. (1.4) arriving 
at the equation
$$
-E+V(r)-{1\over 4r^{2}}=k''(r)+(k'(r))^{2},
\eqno (4.5)
$$
because $l=0$ in the ground state. The above equations lead 
therefore to an algebraic equation where we equate coefficients
of $r^{p}$, for all $p=-4,-3,-2,-1,0$. Hence we find
$$
a^{2}=A,
\eqno (4.6)
$$
$$
2a(1-c)=B,
\eqno (4.7)
$$
$$
c(c-1)-2ab=C-{1\over 4},
\eqno (4.8)
$$
$$
2bc=D,
\eqno (4.9)
$$
$$
b^{2}=-E.
\eqno (4.10)
$$
This system is solved by
$$
a=-\sqrt{A},
\eqno (4.11)
$$
$$
c=1-{B\over 2a}=1+{B\over 2\sqrt{A}},
\eqno (4.12)
$$
$$
b={D\over 2c}={D\over 2 \left(1+{B\over 2 \sqrt{A}}\right)},
\eqno (4.13)
$$
$$
E=-b^{2}=-{D^{2}\over 4 \left(1+{B\over 2 \sqrt{A}}\right)^{2}}.
\eqno (4.14)
$$
Note that, while (4.11) is consistent with a negative value of $a$
as specified in (4.3), negative values of $b$ are only obtained 
if $D<0$ and $c>0$ or if $D>0$ and $c<0$. Moreover, after defining
$$
\mu \equiv {B\over 2 \sqrt{A}},
\eqno (4.15)
$$
Eq. (4.8) can be expressed in the convenient form
$$
C={1\over 4}+\mu(1+\mu)+{D \sqrt{A}\over (1+\mu)}.
\eqno (4.16)
$$
For the evaluation of excited states, one has to write the ansatz
for $y(r)$ in the form of a product, and the resulting analysis
is much harder. For this purpose, a separate paper is in order,
which goes beyond the aims of the present work.
\vskip 0.3cm
\leftline {\bf 5. CONCLUDING REMARKS}
\vskip 0.3cm
\noindent
The first original result of our paper is given by the formulae 
(2.11)--(2.14) and (3.1)--(3.6) for the
solutions of the stationary Schr\"{o}dinger equation
when a central potential of the form (2.1) is considered in
$q$ spatial dimensions, if $\beta$ is even and larger than $4$. 
If $\beta=4$, we recover instead the standard result, according 
to which the regular eigenfunction behaves as 
$r e^{-\sqrt{\alpha}r^{-1}}$ in the neighbourhood of the
origin. Although much work had been done in the literature on
similar problems [23], an investigation as the one we have proposed
was missing to our knowledge.

When the parameter $\beta$ is odd, or for non-integer values of
$\beta$ larger than 2, the equations of Sec. 2 remain valid,
but it is no longer possible to develop a recursive algorithm as
we have done in Sec. 3. 
At the mathematical level, the issue of self-adjoint extensions of
our Schr\"{o}dinger operators deserves careful consideration
as well (cf. [24]).

Last, in Sec. 4, the complete form of the ground-state wave
function has been obtained in two dimensions when the potential
takes the form (4.1). The energy eigenvalue is then given by
Eq. (4.14), provided that the conditions (4.11)--(4.13) 
and (4.16) hold.
\vskip 0.3cm
\noindent
{\bf Acknowledgments.}
This work was supported by the National
Natural Science Foundation of China and Grant No. LWTZ-1298 from
the Chinese Academy of Sciences. The work of GE has been partially
supported by PRIN97 `Sintesi'.
\vskip 0.3cm
\leftline {\bf REFERENCES}
\vskip 0.3cm
\noindent
\item {[1]} 
G. C. Maitland, M. Rigby, E. B. Smith and W. A. Wakeham,  
{\it Intermolecular Forces} (Oxford University Press, Oxford, 1987). 
\item {[2]} 
R. J. LeRoy and W. Lam, {\it Chem. Phys. Lett.} {\bf 71}, 544 (1970);
R. J. LeRoy and R. B. Bernstein, {\it J. Chem. Phys.} 
{\bf 52}, 3869 (1970).
\item {[3]} 
E. Vogt and G. H. Wannier, {\it Phys. Rev.} {\bf 95}, 1190 (1954).
\item {[4]} 
L. D. Landau and E. M. Lifshitz, {\it Quantum Mechanics}, 
Vol. 3, 3rd Ed. (Pergamon Press, Oxford, 1977);
D. R. Bates and I. Esterman, {\it Advances in Atomic and Molecular Physics},
Vol. 6 (Academic, New York, 1970). 
\item {[5]} 
B. H. Bransden and C. J. Joachain,  
{\it Physics of Atoms and Molecules} (Longman, London, 1983). 
\item {[6]} 
S. \"{O}zcelik and M. Simsek, {\it Phys. Lett. A} 
{\bf 152}, 145 (1991). 
\item {[7]} 
R. S. Kaushal and D. Parashar, {\it Phys. Lett. A} 
{\bf 170}, 335 (1992).
\item {[8]} 
R. S. Kaushal, {\it Ann. Phys. (N.Y.)} {\bf 206}, 90 (1991).
\item {[9]} 
S. H. Dong and Z. Q. Ma, {\it J. Phys. A} {\bf 31}, 9855 (1998).
\item {[10]} 
S. H. Dong and Z. Q. Ma, `Exact solutions of the
Schr\"{o}dinger Equation with the Sextic Potential in Two Dimensions'
(submitted to {\it J. Phys. A}). 
\item {[11]} 
S. H. Dong and Z. Q. Ma, `An exact solution
of the Schr\"{o}dinger equation with the octic
potential in two dimensions' (submitted to {\it J. Phys. A}). 
\item {[12]}
V. de Alfaro and T. Regge, {\it Potential Scattering}
(North Holland, Amsterdam, 1965).
\item {[13]}
S. Fubini and R. Stroffolini, {\it Nuovo Cimento} 
{\bf 37}, 1812 (1965).
\item {[14]}
F. Calogero, {\it Variable Phase Approach to Potential
Scattering} (Academic, New York, 1967). 
\item {[15]}
R. G. Newton, {\it Scattering Theory of Waves and Particles}
(McGraw Hill, New York, 1967).
\item {[16]}
W. M. Frank, D. J. Land and R. M. Spector, {\it Rev. Mod. Phys.}
{\bf 43}, 36 (1971).
\item {[17]}
R. Stroffolini, {\it Nuovo Cimento A} {\bf 2}, 793 (1971).
\item {[18]}
G. Esposito, {\it J. Phys. A} {\bf 31}, 9493 (1998).
\item {[19]}
A. O. Barut, {\it J. Math. Phys.} {\bf 21}, 568 (1980).
\item {[20]}
A. V. Turbiner, {\it Commun. Math. Phys.} {\bf 118}, 467 (1988).
\item {[21]}
G. Esposito, {\it Found. Phys. Lett.} {\bf 11}, 535 (1998).
\item {[22]}
M. Abramowitz and I. A. Stegun, {\it Handbook of Mathematical
Functions} (Dover, New York, 1964).
\item {[23]}
E. M. Harrell, {\it Ann. Phys. (N.Y.)} {\bf 105}, 379 (1977).
\item {[24]}
W. Bulla and F. Gesztesy, {\it J. Math. Phys.}
{\bf 26}, 2520 (1985).

\bye